\documentstyle{l-aa}
\begin{document}

\thesaurus{13.07.1, 03.13.5}

\title{Early detection of the Optical Transient following the Gamma--Ray 
Burst GRB 970228$^{\star}$}
\thanks{Based on observations obtained at the Osservatorio Astronomico di 
Loiano, Italy}

\author{A. Guarnieri\inst{1}, C. Bartolini\inst{1}, N. Masetti\inst{1},
A. Piccioni\inst{1}, E. Costa\inst{2}, M. Feroci\inst{2}, F. Frontera\inst{3},
D. Dal Fiume\inst{4}, L. Nicastro\inst{4}, E. Palazzi\inst{4},
A.J. Castro--Tirado\inst{5} \& J. Gorosabel\inst{5}}

\institute{(1) Dipartimento di Astronomia, Universit\`a di Bologna, 
via Zamboni, 33 I-40126 Bologna, Italy\\
(2) Istituto di Astrofisica Spaziale, CNR, Frascati\\
(3) Dipartimento di Fisica, Universit\`a di Ferrara\\
(4) Istituto Tecnologie e Studio Radiazioni Extraterrestri, CNR, Bologna\\
(5) Laboratorio de Astrof\'{\i}sica Espacial y F\'{\i}sica Fundamental,
INTA, Madrid}

\offprints{A. Guarnieri, adriano@astbo3.bo.astro.it}

\date{Received 12 May 1997; accepted 26 June 1997}

\maketitle
\markboth{A. Guarnieri et al.: The Optical Transient of GRB 970228}{}

\begin{abstract}

The optical counterpart of the Gamma--Ray Burst GRB 970228, discovered by
Groot et al. (1997a), is also detected in the $B$ and $R$ frames obtained 
$\sim$4 hours earlier at the Bologna Observatory. Our observations indicate a 
very likely rise of the optical emission within these 4 hours. The $R$ 
luminosity of the transient at maximum was about 15 times that of an 
underlying extended object. Follow--up data show that 
the maximum optical emission was delayed of not less than 0.7 
days with respect to the $\gamma$--ray peak and that no new big flares were 
seen after the main one. The optical transient became significantly redder once
it has reverted to quiescence.


\end{abstract}

\section{Introduction}

Gamma--Ray Bursts are strong high--energy flashes lasting a few seconds
on average and showing an isotropic sky distribution (Fishman \& Meegan 1995).
Their emission in  
the other parts of electromagnetic spectrum is very elusive. Their distance is 
currently unknown, since no optical or radio counterpart has been detected so 
far. This situation seems to be rapidly changing thanks to the reduced error 
boxes now available. Therefore, it is of paramount importance to observe the 
error boxes of these bursts in the optical bands as soon as possible after the 
event, in order to catch the afterglow proposed by several theories. GRB 970228
offered for the first time the possibility of searching for an optical event 
shortly after a burst and to follow its evolution.
Indeed an Optical Transient (OT) was suggested (Groot et al. 
1997a, van Paradijs et al. 1997) as the optical counterpart of this Gamma--Ray 
Burst. 

\bigskip
GRB 970228 was detected on February 28, 1997, by the X--ray satellite BeppoSAX 
(Costa et al. 1997a) 
as a four--peaked $\gamma$--ray burst at $\alpha=5^{\rm h} 01^{\rm m} 
57^{\rm s}$, $\delta=11^{\circ} 46'.4$ (equinox 2000.0; error box radius = 3'),
with total duration 80 seconds.
Eight hours after the event, Costa et al. (1997b) observed a transient X--ray 
source, now labelled as SAX J0501.7+1146, located at the edge of the GRB 970228 
error box. 
Palmer et al. (1997), Matz et al. (1997) and Liang et al. (1997) analyzed the 
spectral 
and the decay behaviours of both GRB 970228 and SAX J0501.7+1146; in 
particular, it was found that the $\gamma$--ray spectrum was consistent with 
that of a classical burst.
Preliminary comparison with the Digital Sky 
Survey showed no brightening of sources to $V=19$ (Groot et al. 1997b), $B=19$ 
and $R=20$ (Guarnieri et al. 1997). Hurley et 
al. (1997), using the Ulysses satellite data, reduced the error box of GRB 
970228 and found that the SAX J0501.7+1146 error box overlapped partially that 
of the $\gamma$--ray burst. Within the intersection of the former error
boxes Groot et al. (1997a) discovered an OT, which they indicated
as probably related to GRB 970228. A thorough presentation of this result
was given by van Paradijs et al. (1997).

\begin{table}
\caption[]{Measured $B$ and $R$ magnitudes of the OT, possible 
counterpart of GRB 970228, plus the nearby red star and the extended object. 
The dates are computed at the times of mid--exposure}
\begin{center}
\begin{tabular}{cccc}
\noalign{\smallskip}
\hline
Day of 1997 (UT) & Band & Magnitude & Exposure time \\
at mid--exposure & & & (seconds) \\
\noalign{\smallskip}
\hline
\noalign{\smallskip}
\multicolumn{1}{l}{Feb. 28.827} & $R$ & $21.1\pm0.2$ & 1800 \\
\multicolumn{1}{l}{Feb. 28.861} & $B$ & $22.3\pm0.3$ & 3600 \\
\multicolumn{1}{l}{Mar. 1.791} & $R$ & $>$21.4 & 1600 \\
\multicolumn{1}{l}{Mar. 3.764} & $R$ & $22.3\pm0.5$ & 1800 \\
\multicolumn{1}{l}{Mar. 4.791} & $R$ & $22.2\pm0.5$ & 1800 \\
\multicolumn{1}{l}{Mar. 4.853} & $B$ & $>$22.5 & 3600 \\
\multicolumn{1}{l}{Mar. 5.858} & $R$ & $>$22.2 & 1800 \\
\multicolumn{1}{l}{Mar. 12.802} & $R$ & $22.6\pm0.5$ & 1800 \\
\multicolumn{1}{l}{Mar. 12.843} & $B$ & $>$22.4 & 3600 \\
\multicolumn{1}{l}{Mar. 13.817} & $B$ & $>$21.5 & 3600 \\
\multicolumn{1}{l}{Mar. 13.852} & $R$ & $>$22.3 & 1800 \\
\multicolumn{1}{l}{Mar. 18.815} & $R$ & $>$20.6 & 2400 \\
\noalign{\smallskip}
\hline
\end{tabular}
\end{center}
\end{table}

In this paper we present the optical photometric analysis of the error box of 
GRB 970228, started $\sim$17 hours after the gamma--ray event, from February 
28 to March 18, 1997.
Section 2 will present the data along with the reduction and calibration 
techniques, while Sect. 3 will discuss the results. Finally, Sect. 4 
will draw the conclusions.

\section{Observations and analysis}

We started an optical observational campaign on the GRB 970228 error box
in order to search for possible optical counterparts of the $\gamma$--ray event,
beginning 15.5 hours after the burst (Guarnieri et al. 1997). Owing to the 
large initial error box (15' in radius; Costa \& Frontera 1997) it was not 
possible to cover it with a single CCD frame; we succeeded to locate the 
``right" field $\sim$16.5 hours after the event, i.e. on February 28.816 UT.

The frames were obtained with the 1.5-meter telescope of the Bologna
University equipped with the BFOSC instrumentation (Merighi et al. 1994), 
which allows fast switching from the spectrographic to the imaging mode. 
Images have a scale of about 0.5 arcsec pixel$^{-1}$. The field was 
observed on February 28 and on March 1, 3, 4, 5, 12, 13 and 18, 1997.
$R$ and $B$ filters were employed, with exposure times spanning from
30 to 60 minutes. The limiting magnitude strongly depended on the 
seeing and on the sky conditions, which were fairly good on February 28, 
March 3, 4 and 12  (typical FWHM of the PSF: 4 pixels).
On March 1, 5, 13 and 18 the bad seeing and poor sky conditions implied
frames with a low signal--to--noise ratio; therefore, they gave us only lower 
limits for the magnitudes of the object.

$B$, $V$ and $R$ images of the Selected Area 101 (Landolt 1992) were also 
obtained on March 4 in order to calibrate the field of GRB 970228.
After the standard cleaning procedure for bias and flat field, the frames
were processed with the DAOPHOT II package (Stetson 1987) and the 
{\sl ALLSTAR} procedure inside MIDAS. We used simple aperture photometry when
the objects were too faint to be detected with DAOPHOT II.
Then, the fields were calibrated with the standard stars quoted above.

We detected an OT
at the intersection of the error boxes of GRB 970228 and SAX J0501.7+1146
with coordinates $\alpha=5^{\rm h} 01^{\rm m} 47^{\rm s}$, 
$\delta=11^{\circ} 46' 55"$ (equinox 2000.0; error: $\pm$5").
Its position makes it practically coincident with the proposed 
optical counterpart of GRB 970228. Its magnitude changed from $R=21.1\pm0.2$ 
and $B=22.3\pm0.3$ on February 28.8 UT to $R=22.3\pm0.5$ on March 3.8 UT and 
$B>22.5$ on March 4.8 UT. In the days following March 
3.8 the $R$ magnitude of the object seemed to remain more or less constant. 
Table 1 shows the measured $B$ and $R$ magnitudes.
The reported values are the integrated magnitudes of the OT 
plus a nearby star and an extended object (van Paradijs et al. 1997).
The latter is possibly
the host of the OT, 0".2 away (van Paradijs et al. 1997, Groot
et al. 1997c) and with $R=24$ (Groot et al. 1997c, Metzger et al. 1997a), 
whereas the nearby object is an early M--type star (van Paradijs et al. 1997, 
Groot et al. 1997c) or a mid K-type star (Tonry et al. 1997), located 2".7 
away, constant in brightness with $R=22.4$ 
(Metzger et al. 1997a) and unrelated to the OT (Groot et al. 1997c). Due to the
seeing conditions, we were not able to separate the OT from the star.

In our frames the red star is more likely responsible for the observed residual
emission already since March 3. Actually, the $R$ magnitude of the object 
after that day, within the errors, is the same as the red star located 
near the extended source as indicated by Metzger et al. (1997a). This 
means that the main optical effects of the $\gamma$--ray explosion (rise and
first decay phase) developed
before March 3.8.
Indeed, observations made on March 6.3, 11.2 (Metzger et al. 1997a), 9.9 and 
13.0 (Groot et al. 1997c) show that the underlying extended object is constant 
in brightness.
Therefore, within the accuracy of the measurements, the time span of 3.6 days 
is the upper limit to the duration of the brightest phase of the point--like 
OT, as Table 1 shows. 
Our observations are consistent with the
fading object found by Groot et al. (1997a) and van Paradijs et al. (1997)
with $V=21.3$ and $I=20.6$ on March 1.0 UT and $V>23.6$ and $I>22.2$ on 
March 8.9 UT. According to van Paradijs et al. (1997) and Metzger et al. 
(1997a), it is $\sim$0".2 from a quiescent object, which appears to be extended
and therefore likely to be a galaxy which could be associated with the X--ray 
transient and, presumably, with the GRB.

\begin{table*}
\caption{X--ray, $B$, $V$ (Johnson), $R$ and $I$ (Cousins) fluxes of the event.
 $B$ and 
$R$ values were computed by subtracting from the data of Table 1 in the days 
from February 28 to March 3 the fluxes of the extended object and of the nearby 
star. Effective wavelengths are reported. We adopted the band widths published
by Fukugita et al. (1995). All fluxes are in units of 10$^{-15}$ erg cm$^{-2}$ 
s$^{-1}$. Data between parentheses are inferred from our interpolations. See 
text for further 
details}
\begin{center}
\begin{tabular}{ccccccc}
\noalign{\smallskip}
\hline
Day of 1997 & X--ray flux & $B$ flux & $V$ flux & $R$ flux & $I$ flux & 
Reference \\
 UT & 0.5 -- 10 keV & 
%$\Delta \lambda=1008$ \AA & $\Delta \lambda=827$ \AA & $\Delta \lambda=1568$ 
%\AA & $\Delta \lambda=1542$ \AA \\
$ \lambda=4448$ \AA & $\lambda=5505$ \AA & $\lambda=6588$ \AA & 
 $\lambda=8060$ \AA & number \\
\noalign{\smallskip}
\hline
\noalign{\smallskip}
\multicolumn{1}{l}{Feb. 28.46} & 4000 & --- & --- & --- & --- & 1 \\
\multicolumn{1}{l}{Feb. 28.827} & --- & $5.7\pm2.5$ & ($4.4\pm2.4$) & 
$7.7\pm2.4$ & --- & 2 \\
\multicolumn{1}{l}{Feb. 28.99} & --- & --- & $9.0\pm0.9$ & ($14.1\pm1.9$) & 
$9.8\pm1.0$ & 2,3,4 \\
\multicolumn{1}{l}{Mar. 1.79} & --- & --- & --- & $<$5.8 & --- & 2 \\
\multicolumn{1}{l}{Mar. 3.73} & 200 & --- & --- & --- & --- & 1 \\
\multicolumn{1}{l}{Mar. 3.76} & --- & --- & --- & $<$2.4 & --- & 2 \\
\multicolumn{1}{l}{Mar. 4.86} & --- & --- & $<$0.6 & --- & --- & 4 \\
\multicolumn{1}{l}{Mar. 8.9} & --- & --- & $<$1.1 & --- & $<$2.3 & 3,4 \\
\multicolumn{1}{l}{Mar. 26.4} & --- & --- & $0.11\pm0.01$ & ($0.27\pm0.03$) & 
$0.36\pm0.04$ & 2,5 \\
\multicolumn{1}{l}{Apr. 7.2} & --- & --- & $0.08\pm0.01$ & ($0.20\pm0.03$) 
& $0.25\pm0.02$ & 2,5 \\
\noalign{\smallskip}
\hline
\noalign{\smallskip}
\multicolumn{7}{l}{Refs. --- 1. Costa et al. 1997b; 2. this work; 
3. Groot et al. 1997a; 4. van Paradijs et al. 1997; 5. Sahu et al. 1997.}\\
\noalign{\smallskip}
\hline
\end{tabular}
\end{center}
\end{table*}

We could now subtract from the total fluxes of the optical event those of the 
M--type star and of the extended object. Since the constant red star and the 
underlying extended object were always undetectable in our $B$ band photometry, 
we can assume that the $B$ magnitude on February 28.8 is representative 
of the transient event. We obtain $B$ = 22.4 and $R$ = 21.6, implying 
$B-R$ = 0.8 as the color index of the OT $\sim$17 hours after 
the $\gamma$--ray event. On March 3.8 we deduce that the OT was
fainter than $R$ = 23. Van Paradijs et al. (1997) report a fading $\Delta V
>2.9$ mag in the time interval March 1.0--March 4.9. 
Pedichini et al. (1997) observed a fading of 2.7 mag between 
February 28.8 and March 4.8.

\section{Discussion}

The combination of Bologna ($B=22.4\pm0.4$, $R=21.6\pm0.3$; this work) and La
Palma ($V=21.3\pm 0.1$, $I=20.6\pm0.1$; van Paradijs et al. 1997) data taken 
on February 28 could provide very important informations to understand the 
nature of this transient, so far unique in the optical bands. 
These observations are not simultaneous, therefore they can give 
some insights into the problem of the light variation and into the related one
of the energy distribution. We consider here three different hypotheses:
a) the brightness was constant during the four hours between
the two sets of observations, one obtains $B-V=1.1$, $V-R=-0.3$, and $R-I=1.0$;
such color indices are not consistent with the spectral energy distribution 
of any known astrophysical object and imply strong variations with the 
wavelength. b) a fading during the time span between Bologna and La Palma 
observations; in this case the above result would be strengthened.
c) the possibility of an increasing emission. In the latter case, we can 
interpolate the 
$V$ spectral flux density from $B$ and $R$ values at the time of the Bologna 
observations.  We assume tentatively a linear flux density $\Delta F$/$\Delta 
\lambda$ distribution. 
The conversion from magnitudes to fluxes was done using Table 9 by 
Fukugita et al. (1995).
We obtain $5.9\times10^{-18}$ erg cm$^{-2}$ s$^{-1}$ \AA$^{-1}$,
corresponding to $V=22.0$. In the same way, from La Palma $V$ and 
$I$ magnitudes we derive a $R$ spectral flux density of $9.0\times10^{-18}$ 
erg cm$^{-2}$ s$^{-1}$ \AA$^{-1}$, which gives $R=20.9$.
Thus, the total $R$ flux at maximum was not less than 15 times that of the 
extended object.
These figures imply a flux increase of 1.9 times in $V$ and 
$R$ between the two sets of observations, corresponding to a variation of 
$-$0.7 mag in both bands, with a mean rate of $\sim$ $-$0.2 mag 
hr$^{-1}$. The conclusion seems to be inescapable: either the OT has a 
very bizarre spectrum, or if it has a more normal spectral distribution 
it must display an increase of brightness between Bologna and La Palma 
observations. In the framework of the foregoing hypothesis, this result is 
significant at a 2$\sigma$ confidence level.
Being cautious for the observational error, we suggest
that the optical luminosity increased at least until March 1.0 UT. Indeed
a quick computation shows that, in order to have constancy or a 
decrease in brightness with a 3$\sigma$ confidence level between Bologna and 
La Palma data, we should have observed the OT at least at $R=20.5$ on 
February 28.8; this value is outside the 3$\sigma$ interval centered on the 
observed magnitude $R=21.6\pm0.3$. We then conclude that the hypothesis of a 
non--increase in brightness can be rejected with a confidence level of almost
4$\sigma$.

We can now refine our first--approximation figures. On February 28, the 
Bologna $B$ value was acquired about 50 minutes later than the $R$ one, 
therefore we might infer that the $B$ magnitude of the OT at 
the time of the $R$ frame was actually brighter by $\sim$0.2 mag. This implies 
a simultaneous $B-R$ color index of $\sim$1.0; correspondingly, one has
$B=22.6$ and $V=22.1$ (interpolated at the same time).
Bearing this in mind, it results that in a time span of about 4 hours the $V$ 
magnitude of the OT decreased of 0.8 mag; this corresponds
to a flux increase of a factor $\sim$2.1. 

HST data (Sahu et al. 1997) are particularly relevant for understanding the 
mid--term behaviour of the OT. Therefore, by using the method described
above, we computed the $R$ flux densities and magnitudes; we obtained $R=25.2$
for March 26 and $R=25.6$ for April 7.
From these data it is evident a substantial reddening of the OT, which changed 
from $V-R=0.4$ and $V-I=0.7$ on February 28.99 to $V-R=0.9$ and 
$V-I=1.9$ on March 26.4.
The fluxes corresponding to these interpolations, together with 
those of other relevant $B$, $V$, $R$ and $I$ measurements and with the 
fluxes of the X--ray transient source SAX J0501.7+1146 given by Costa et al. 
(1997b) are reported in Table 2.

\smallskip
In spite of the poor quality of our observation of March 1.8 UT we can state
that, on the basis of a comparison with other objects in the frame, at that 
time the OT had faded below the level of our first detection.
Therefore we can fix the time delay ${\tau_d}$ between the $\gamma$--ray  
event and optical maximum in the range
$0^{\rm d}.71<{\tau_d}<1^{\rm d}.67$. Correspondingly, the duration 
${\tau_f}$ of the first fading phase is ${\tau_f} < 2^{\rm d}.93$. During this
time the fading rate is $\approx$1 mag day$^{-1}$.
This fixes tight constraints on both the rising and fading optical rates.
The presence of an early phase of rapid fading is implicitly confirmed by
the $R$ band observations of Metzger et al. (1997a,b) who found that the total 
$R$ magnitude has slowly faded by $1.0\pm0.4$ mag in one month 
(March 6--April 6).

If we assume for the $R$ band luminosity a power law decay 
$L_{\rm opt ({\it R})}\propto t^{-\alpha}$, a limit $\alpha>1.1$ can be 
determined by using the $R$ flux (interpolated as before) from the HST data of 
March 26.4 (Sahu et al. 1997); the optical data of the first days require 
$\alpha>1.4$.
An index $\alpha\sim1.4$ is also found for the X--ray decay behaviour. An 
exponential decay law, similar to that of X--ray bursts, results in a 
decay time $<$1$^{\rm d}.3$, to be compared with
$\sim$1$^{\rm d}$ deduced by Palmer et al. (1997) for the X--ray emission. 
The ratio $L_{\rm opt ({\it R})}/L_{\rm X (0.5-10~keV)}$ is 
$\approx$ 4$\times10^{-3}$, if the luminosity values are taken at their 
respective observed maxima (at least 13 hours apart). By scaling the X--ray flux
at the time of optical maximum (with a $1^{\rm d}$ decay law), we obtain
$L_{\rm opt ({\it R})}/L_{\rm X}\approx6\times10^{-3}$. On March 3 X--ray and
optical observations were nearly simultaneous: we can fix an upper limit of 
$\approx$1.2$\times10^{-2}$ to the $L_{\rm opt ({\it R})}/L_{\rm X}$ ratio.

It is interesting to note that Castro--Tirado et al. (1997) did not see anything
noteworthy in the error box of another Gamma--Ray Burst, GRB 970111, just 19 
hours after the $\gamma$--ray event. This indicates a significant difference
in the optical behaviour of GRB 970111 and GRB 970228.

\section{Conclusions}

We can now summarize the main results of this work:

\begin{enumerate}

\item 
our observations in $B$ and $R$ bands, probably the earliest to the GRB 970228 
event, clearly reveal the fading object (Groot et al. 1997a, van Paradijs et 
al. 1997) in the intersection of the combined error boxes of GRB 970228 and 
SAX J0501.7+1146;

\item
they also indicate an increase in luminosity of a factor $\sim$2 on a time 
scale of 4 hours;

\item
the bulk of the optical event (rising and first decay phase) developed in no 
more than $3^{\rm d}.6$;

\item
the maximum of the optical emission from the `fireball' was attained
probably not earlier than $0^{\rm d}.71$ after the $\gamma$--ray burst;

\item
the ratio $L_{\rm opt ({\it R})}/L_{\rm X (0.5-10~keV)}$ at the supposed 
optical maximum is $\approx$6$\times10^{-3}$;

\item
the $R$ luminosity of the OT at maximum is about 15 times that of the 
underlying extended object;

\item
our sampling of the source until March 18 shows that after the first
there were no new big flares;

\item
the color indices significantly reddened during the month after February 28.
\end{enumerate}

In conclusion, if the optical fading object is the counterpart of GRB 970228, 
all this severely constrains the modeling of Gamma--Ray Bursts.

\begin{acknowledgements}

This investigation is supported by the University of 
Bologna (Funds for selected research topics). We acknowledge the use of the 
BFOSC instrument of the Bologna Astronomical Observatory.
We thank M. Tavani, G.M. Beskin and M. Corwin for useful 
comments and suggestions. We also thank A. Bragaglia, P. Focardi, A. Comastri 
and G. Tozzi for having given us part of their observational time. 

\end{acknowledgements}

\bigskip
{\bf Note added in proof}.
The proposed optical counterpart of GRB 970508 
confirms the existence of a
significant time delay between the $\gamma$--burst and the maximum of optical
brightness, as shown in this work.

\end{document}